\documentclass[a4paper,reqno]{amsart}
\usepackage{color}
\usepackage{enumerate}
\usepackage{hyperref}
\usepackage{marginnote}
\usepackage{amsmath, amsthm, amssymb, comment, array,accents}
\usepackage{tikz, pgfplots,xcolor}
\usetikzlibrary{decorations.markings, shapes.misc, positioning}
\usetikzlibrary{angles, arrows, automata, backgrounds, calc, calendar, chains, decorations.pathmorphing, decorations.pathreplacing, intersections,  matrix, mindmap, patterns, quotes, shapes}
\usepackage[latin1]{inputenc}
\usepackage{booktabs,bm}
\usepackage{url,fancyref,hyperref}
\hypersetup{colorlinks,citecolor=red,urlcolor=blue,linkcolor=black,filecolor=black}
\usepackage[toc]{appendix}
\usepackage{marginnote}
\usepackage[inline,shortlabels]{enumitem}

\usepackage{longtable}
\newcolumntype{L}{>{$}l<{$}}
\newtheoremstyle{mytheoremstyle} 
    {\topsep}                    
    {\topsep}                    
    {\itshape}                   
    {}                           
    {\bfseries}                   
    {:}                          
    {.5em}                       
    {}  

\newtheoremstyle{mydefinitionstyle} 
    {\partopsep}                    
    {\partopsep}                    
    {\normalfont}                   
    {}                           
    {\bfseries}                   
    {:}                          
    {.5em}                       
    {}  

\makeatletter
\def\th@plain{%
  \thm@notefont{}
  \itshape 
}
\def\th@mydefinitionstyle{%
  \thm@notefont{}
  \normalfont 
}
\makeatother
\let\oldsqrt\sqrt
\def\sqrt{\mathpalette\DHLhksqrt}
\def\DHLhksqrt#1#2{%
\setbox0=\hbox{$#1\oldsqrt{#2\,}$}\dimen0=\ht0
\advance\dimen0-0.2\ht0
\setbox2=\hbox{\vrule height\ht0 depth -\dimen0}%
{\box0\lower0.4pt\box2}}

\theoremstyle{mytheoremstyle}

\theoremstyle{mydefinitionstyle}

\newcommand{\be}{\begin{equation}}
\newcommand{\ee}{\end{equation}}
\newcommand{\bal}{\begin{align*}}
\newcommand{\eal}{\end{align*}}
\newcommand{\ba}{\begin{array}}
\newcommand{\ea}{\end{array}}
\newcommand{\bmx}{\begin{bmatrix}}
\newcommand{\emx}{\end{bmatrix}}
\newcommand{\ben}{\begin{equation*}}
\newcommand{\een}{\end{equation*}}
\newcommand{\beq}{\begin{eqnarray}}
\newcommand{\eeq}{\end{eqnarray}}
\newcommand{\beqn}{\begin{eqnarray*}}
\newcommand{\eeqn}{\end{eqnarray*}}
\newcommand{\bseq}{\begin{subequations}}
\newcommand{\eseq}{\end{subequations}}
\newcommand{\bmat}{\begin{pmatrix}}
\newcommand{\emat}{\end{pmatrix}}
\newcommand{\bem}{\begin{enumerate}}
\newcommand{\eem}{\end{enumerate}}
\newcommand{\bd}{\begin{description}}
\newcommand{\ed}{\end{description}}
\newcommand{\bl}{\begin{itemize}}
\newcommand{\el}{\end{itemize}}
\newcommand{\bc}{\begin{cases}}
\newcommand{\ec}{\end{cases}}

\renewcommand{\tilde}{\widetilde}
\newcommand{\al}{\alpha}

\newcommand{\gm}{\gamma}

\newcommand{\de}{\delta}

\newcommand{\om}{\omega}
\newcommand{\Om}{\Omega}
\newcommand{\ka}{\kappa}
\newcommand{\tht}{\theta}

\newcommand{\Gm}{\Gamma}

\newcommand{\omb}{\overline{\omega}}

\newcommand{\omt}{\tilde{\om}}

\newcommand{\Omb}{\overline{\Om}}
\newcommand{\phit}{\tilde{\phi}}

\newcommand{\Phib}{\overline{\Phi}}

\newcommand{\vh}{\hat{v}}

\newcommand{\wc}{\accentset{\circ}{w}}

\newcommand{\Eb}{\overline{E}}

\newcommand{\zb}{\overline{z}}

\newcommand{\nn}{\nonumber}

\newcommand{\ord}{\mathcal{O}}

\newcommand{\pv}{\text{Painlev\'e }}

\newcommand{\pI}{\text{P$_{\text{I}}$}}
\newcommand{\pII}{\text{P$_{\text{II}}$}}
\newcommand{\pIII}{\text{P$_{\text{III}}$}}
\newcommand{\pIV}{\text{P$_{\text{IV}}$}}
\newcommand{\pV}{\text{P$_{\text{V}}$}}
\newcommand{\pVI}{\text{P$_{\text{VI}}$}}

\newcommand{\pa}{\partial}

\newcommand{\Ra}{\Rightarrow}

\newcommand{\bs}{\backslash}
\newcommand{\cmmnt}[1]{}

\theoremstyle{definition}
\newtheorem{definition}{Definition}[section]
\theoremstyle{remark}
\newtheorem{rem}[definition]{Remark}
\theoremstyle{theorem}
\newtheorem{thm}[definition]{Theorem}
\newcommand{\pJ}{\text{P$_{\text{J}}$}}
\numberwithin{equation}{section}

\allowdisplaybreaks
\begin{document}
\title[Asymptotics of $\pI-\pV$]{Asymptotic Behaviours Given by Elliptic Functions in $\pI$--$\pV$}
\author{Nalini Joshi}
\address{School of Mathematics and Statistics F07, The University of Sydney, NSW 2006, Australia}
\email{nalini.joshi@sydney.edu.au}
\thanks{This research was supported by an Australian Laureate Fellowship \# FL 120100094 from the Australian Research Council. }
\author{Elynor Liu}
\email{e.liu@maths.usyd.edu.au}
\address{School of Mathematics and Statistics F07, The University of Sydney, NSW 2006 Australia}
\thanks{EL's research was supported by a postgraduate research award from the University of Sydney.}
\date{}

\begin{abstract}
Following the study of complex elliptic-function-type asymptotic behaviours of the \pv equations by Boutroux and Joshi and Kruskal for $\pI$ and $\pII$, we provide new results for elliptic-function-type behaviours admitted by $\pIII$, $\pIV$, and $\pV$, in the limit as the independent variable $z$ approaches infinity. We show how the Hamiltonian $E_{\rm J}$ of each equation $\rm P_{\rm J}$, $\rm J=I, \ldots , V$, varies across a local period parallelogram of the leading-order behaviour, by applying the method of averaging in the complex $z$-plane. Surprisingly, our results show that all the equations $\pI-\pV$ share the same modulation of $E$ to the first two orders.
\end{abstract}
\maketitle

\section{Introduction}\label{sec:intro}
Interest in the Painlev\'e equations has grown since they were recognised as universal models in mathematical physics, in fields such as random matrix theory \cite{Deift2007, For2003, TW1994}, quantum gravity \cite{Gross1990a}, quantum field theory \cite{CTW1981}, nonlinear optics \cite{GJ1989}, general relativity \cite{Mac1983} and statistical mechanics \cite{Levi1992}. Critical solutions of interest in such applications were first identified through asymptotic analysis \cite{WMTB76,MTW77} (see also \cite{FIKN2006}), usually as the independent variable approaches a fixed singular point at infinity. However, the asymptotic analysis of the Painlev\'e equations remains incomplete in the literature, particularly for the third, fourth and fifth Painlev\'e equations in the complex plane. In this paper, we provide new results describing generic leading order behaviours of these equations.

To make the asymptotic analysis more explicit, we write the first five Painlev\'e equations in the following form (known as Boutroux form for $\pI-\pII$ \cite{Bout13})
\begin{align}
\pI: u_{zz}&=6u^2-1-\,\frac{u_z}{z}+\frac{4u}{25 z^2},\label{p1u}\\
\pII: u_{zz}&=2u^3-u-\,\frac{\alpha+u_z}{z}+\frac{u}{9 z^2},\label{p2u}\\
\pIII:  u_{zz}&=\frac{u_z^2}{u} +\frac \delta u +\gm\, u^3 + \frac{\al\, u^2+\beta}{z} - \frac{u_z}z,\label{p3u}\\
\pIV:  u_{zz}&=\frac{u_z^2}{2\,u}+2\,u+4\,u^2+\frac{3}{2}u^3-\frac{\al\,u}{z}-\frac{u_z}{z}+\frac{\beta}{4\,z^2\,u}+\frac{u}{8\, z^2},\label{p4u}
\\
\pV:  u_{zz}&=\left(\frac{1}{2u}+\frac{1}{u-1}\right)u_z^2-\frac {u_z}{z} +\frac{(u-1)^2}{z^2}\left(\al\, u+\frac{\beta}{u}\right)+\gm\frac u z+\de\, u\frac{u+1}{u-1}.\label{p5u}
\end{align}
Note that $\pIII$ and $\pV$ are in the original form \cite{G09}, and $\pIV$ is a transformed version of its conventional form. (See Appendix \ref{A} for the transformations required for $\pI$, $\pII$ and $\pIV$.)

In the limit $|z|\to\infty$, the above equations become autonomous and these limiting equations have first integrals. This observation leads us to the following definitions of slowly varying quantities that play a crucial role in our analysis of $\pI-\pV$:
\begin{subequations}\label{eq:Ham}
\begin{align}
E_{\rm I}:=&\frac12\left(u_z^2-4u^3+2u\right),\\
E_{\rm II}:=&\frac12\left(u_z^2-u^4+u^2\right),\\
E_{\rm III}:=&\frac{1}{2\,u^2}\left(u_z^2-\gm u^4+\de\right),\label{eq:HamIII}\\
E_{\rm IV}:=&\frac{1}{2\,u}\left(u_z^2-u^4-4\,u^3-4\,u^2\right),\\
E_{\rm V}:=&\frac{1}{2\,u\,(u-1)^2}\left(u_z^2+2\,\de\,u^2\right).
\end{align}
\end{subequations}
Note that these may be considered as Hamiltonians and are conjugate to the Hamiltonians in the literature \cite{O80}. We rewrite equations \eqref{eq:Ham} as
\begin{equation}\label{e:uprimeJ}
u_z^2=:P_{\rm J}(u; E_{\rm J}),
\end{equation}
for $\rm J=\rm I,\ldots \rm V$, and define the corresponding functionals $\om^{(\rm J)}_k$, $\omt^{(\rm J)}_k$ by
\beq\label{periodints}
\om^{(\rm J)}_k&=&\oint_{C_k} \frac{du}{\sqrt{P_{\rm J}(u; E_{\rm J})}}=\frac{d}{dE_{\rm J}}\omt^{(\rm J)}_k,
\eeq
on a universal covering space for $u\in\mathbb C$, where $C_k$, $k=1, 2$ are independent closed contours enclosing a pair of roots of $P_{\rm J}$. The leading-order behaviour of each equation $\pJ$ is given by an elliptic function with periods given by $\{\om^{(\rm J)}_1, \om^{(\rm J)}_2\}$.  

For each $J=\rm I, \dots, V$, the Picard-Fuchs equations for these elliptic integrals take the following form:
\beq\label{pfeq}
\frac{d^2\omt^{(\rm J)}_k}{dE_{\rm J}^2}=F_{\rm J}(E_{\rm J})\omt^{(\rm J)}_k, 
\eeq
where $k=1,2$. The expressions for the coefficients $F_{\rm J}(E_{\rm J})$ are given in Theorem \ref{thm} below (see section \ref{s:main}). 

Such behaviours were first considered by Boutroux \cite{Bout13, Bout14}. For $\pI$ and $\pII$, Joshi and Kruskal \cite{JoshiThesis, JK1988,JK1992} analysed leading-order elliptic-function behaviours in the limit $|z|\to\infty$ for $z\in\mathbb C$ by developing a complex averaging method to obtain the modulation of such behaviours as the angle of approach to infinity changes. We obtain such modulations for the first time for $\pIII$, $\pIV$ and $\pV$ in this paper. 

The prevailing method of asymptotic study for the $\pv$ equations in the literature relies on the Riemann-Hilbert or isomonodromy approach applied to linear equations associated with the $\pv$ equations \cite{FIKN2006}. A review of the application of the isomonodromy method for elliptic-function-type behaviours of ${\rm P}_{\rm I}$ and ${\rm P}_{\rm II}$ can be found in \cite{Kitaev94}. Such studies were motivated by the need to find connections between asymptotic behaviours that hold along certain distinguished directions in the complex plane. In the Boutroux form for ${\rm P}_{\rm J}$ given above, these directions are given by semi-axes lying along the real and imaginary directions for $z\in\mathbb C$. To the best of our knowledge, the Riemann-Hilbert approach has not been used to find the modulation of elliptic-function-type behaviours in the interior of sectors bounded by such rays, i.e., quadrants in $z\in\mathbb C$.  
In the present paper, we focus on the problem in the interior of quadrants and deduce new results for the third, fourth and fifth Painlev\'e equations.  

Our main result is given by Theorem \ref{thm} (see Section \ref{s:main} below). For each ${\rm J}={\rm III}$, $\rm IV$, $\rm V$, we find that the slow modulation of $E_{\rm J}$ is given to leading order by complete elliptic integrals associated to the leading-order elliptic function. It is surprising to find that the equation for the modulation of $E_{\rm J}$, i.e., Equation \eqref{eq:modulation}, turns out to be identical for $\pI$ to $\pV$; see Theorem \ref{thm} below. This unexpected result suggests that the \pv equations share a fundamental asymptotic property, which has not been observed before.

\subsection{Main result}\label{s:main}
 In this section, we state our main result, namely Theorem \ref{thm}. We start with some preliminary notation to describe initial value problems and and remarks to clarify the hypotheses of the theorem.  
  
For each $\rm J=\rm I,\ldots, \rm V$, suppose $0<\epsilon<1$ and $a, \,p\in\mathbb C$ are given numbers such that $|p|>\epsilon$ and for $\rm J={\rm III},\, {\rm IV},\, \rm V$, $\epsilon<|a|$, while for $\rm J=\rm V$, $\epsilon<|a-1|$. 

\begin{definition}\label{def11}
Assume that $\Phi$ lies in an annulus given by $\mathcal D_\epsilon=\{|z|>1/\epsilon\}$. For each ${\rm P}_{\rm J}$, define initial values
\[
u(\Phi)=a, \; u'(\Phi)=p,
\]
where $a\neq 0$ for $\pIII$ and $\pIV$, and $a\neq 0$ or $1$ for $\pV$.  Define the set of numbers $\Phi$, $a$, $p$ satisfying the above conditions to be {\em admissible}.
\end{definition}

\begin{rem}
Note that standard theorems applied to admissible initial data give a unique solution $u(z)$ in a domain containing $\Phi$ in the interior of $\mathcal D_\epsilon$. By the well known fact that solutions are meromorphic in domains of the universal covering space not containing the fixed singularities of ${\rm P}_{\rm J}$, the domain of existence of $u$ can be extended to a large bounded disk punctured at its movable poles  (see, in particular, the description of \lq\lq cheese-like\rq\rq\ regions in \cite{JK1994}). 
\end{rem}
\begin{rem}
Each admissible $a$ and $p$ gives a value of $E_{\rm J}$ at $\Phi$. For each $\rm J$, we can apply the inverse function theorem to the respective Equation \eqref{eq:Ham} in a neighbourhood of $\Phi$, to get two new values of $z$ at which $u$ takes on the value $a$. This result is related to the fact that \eqref{eq:Ham} defines an elliptic curve for each $\rm J$, with $E_{\rm J}$ providing the modulus of the leading-order elliptic function behaviour. To see that these new points can be reached by analytic continuation of $u$, we apply similar arguments to those given in \cite{JK1994}. The separation between each of these points and $\Phi$ is approximately equal to a period of the leading order behaviour $\om_i$, $i=1,2$, defined above by Equation \eqref{periodints}. These results are proved in Theorem \ref{thm} below.  
Let these new points be given by $\Phi+\Om_i$,  $i=1,2$. Each $\Om_i$ is approximated by . 
\end{rem}
We are now in a position to state our main result. 
\begin{thm}\label{thm} 
Given $0<\epsilon<1$,  let $\Phi$, $a$, $p$ be admissible numbers and assume $u(z)$ is a solution of ${\rm P}_{\rm J}$ satisfying $u(\Phi)=a, u'(\Phi)=p$. Define $E_{\rm J}$ for each respective $\rm J=\rm I,\ldots, \rm V$ by Equations \eqref{eq:Ham}. Then for sufficiently large $\Phi$ there exist $\Om_i$ such that
\be\label{eq:modulation}
E_{\rm J}(\Phi+\Om_i^{(\rm J)})-E_{\rm J}(\Phi)=-\frac{\omt_i^{(\rm J)}}{\Phi}+\frac{\om_i^{(\rm J)}\omt_i^{(\rm J)}}{\Phi^2}+\ord\left(\frac{1}{\Phi^3}\right),
\ee
where $i=1,2$. Moreover, $E_{\rm J}$ is bounded along a ray going to infinity with a fixed angle. The corresponding elliptic integrals $\omt_i^{(\rm J)}$ satisfy a second order Picard-Fuchs differential equation $\omt_i^{(\rm J)}{''}=F_{\rm J}(E)\omt_i^{(\rm J)}$, where primes denote differentiation with respect to $E$. The coefficients $F_{\rm J}$ are given by  
\beqn
F_{\rm I}(E)&=&-\frac{15}{4}\frac{1}{(27E^2-2)},\\
F_{\rm II}(E)&=&-\frac32\frac{1}{E\left(8E-1\right)},\\
F_{\rm III}(E)&=&-\frac14\frac{1}{(E^2+\de\gm)},\\
F_{\rm IV}(E)&=&-\frac{6}{E\left(27 E-16\right)},\\
F_{\rm V}(E)&=&-\frac{1}{E\left(4E+\de\right)}.
\eeqn
Furthermore, the distances to the next repeated initial value in the directions of the two local periods are given by
\be
\Om_i^{(\rm J)}=\om_i^{(\rm J)}-\frac{1}{\Phi}\frac{\omt_i^{(\rm J)}\om^{(\rm J)}_{i} {'}}{2}-\frac{\om^{(\rm J)}_{i} {'}}{\Phi}G_{\rm J}( u(\Phi))+\ord(1/\Phi^2),
\ee

where $'=\frac{d}{dE_{\rm J}}$, and
\begin{align*}
G_{\rm I}(u(z))&=-\frac{4}{15\,u_z}(6u^4 - 9u^2 - 6Eu+2),\\
G_{\rm II}(u(z))&=\al\, u-\frac{u^5-2u^3+(1-2E)u}{3u_z},\\
G_{\rm III}(u(z))&=\frac{\al\,u^2-\beta}{u}-\frac{\gm u^4+\de}{uu_z},\\
G_{\rm IV}(u(z))&=-\al\,u-\frac{3u^4+ 18u^3 + 32u^2 + (16-3E)u}{6u_z},\\
G_{\rm V}(u(z))&=-\frac{\gm}{2}\frac{u+1}{u-1}+\de\frac{u(u+1)}{u_z(u-1)}.
\end{align*}
\end{thm}
In the remainder of the paper, we provide a proof of this theorem for each of ${\rm P}_{\rm J}$, $\rm J={\rm III}, {\rm IV}, {\rm V}$.
 
\subsection{Background}
We use standard notation for asymptotic relations in a limit (see \cite{BO1999} for further details). The method of averaging \cite{SVM2007} is used to describe the modulation of the leading-order asymptotic behaviours approaching infinity, which involve (doubly) periodic functions. In this approach, we describe the solution's behaviour by separating it into two parts, where the first leading-order averaged solution is periodic over a long time scale, while the remaining smaller part of the solution fluctuates on a short time scale.  Typically, the method of averaging is applied after introducing a small parameter. In this paper, the small parameter is understood to be given locally by the size of $1/|z|<\epsilon$. 

We note that the method of multiple-scales could also have been used in this context. An extension of the classical multiple-scales method was developed in \cite{JoshiThesis, JK1988} to study the global asymptotic behaviour of generic solutions of $\pI$ and $\pII$. However, there is a technical requirement for the periods of the leading-order solution to be fixed before multiple-scales is applied. This requirement leads to a non-conformal mapping of the complex plane, which means that the equation and its solutions needed to be embedded in a framework of non-analytic functions. For simplicity, to avoid these technicalities, we focus here on the method of averaging to obtain our results. 

\subsection{Outline of the paper} 
The remainder of the paper gives results for each of $\pIII$, $\pIV$, $\pV$. Section \ref{sec:p3} and Appendix \ref{B} provide a detailed calculation and necessary arguments for the proof of Theorem \ref{thm} for $\pIII$. In Section \ref{sec:p4}, we present an outline of the proof for $\pIV$, while that for $\pV$ is presented in Section \ref{sec:p5}. The paper ends with a conclusion and discussion of the main results in Section 5.

\section{The third \pv equation}\label{sec:p3}
In this section we give the proof of Theorem \ref{thm} for $\pIII$, i.e. Equation \eqref{p3u}. In particular, we carry out the averaging method for the Hamiltonian $E_{\rm III}$, find the Picard-Fuchs equations for the related elliptic integrals, and deduce the asymptotic expansion of $\Omega_i^{\rm (III)}$. For simplicity, we drop the label $\rm J=III$ in this section. 

We first make use of a transformation $u=e^{v}$ so that when we do an expansion the first-derivative term is removed from the leading order equation, see Appendix \ref{A2} for derivation. This transformation is defined on the universal covering space $U$ defined by $\exp: U\to C\bs\{0\}$. One can do this transformation at the step of calculating $E(\Phi+\Om_i)-E(\Phi)$, however we rather apply this transformation at the beginning of the analysis to make consistency manifest. Equation \eqref{p3u} becomes
\be\label{p3inv}
v_{zz}=\gm e^{2v} + \de e^{-2v} -\frac{v_z}{z}+\frac{\al e^v+\beta e^{-v}}{z}.
\ee
Integrating equation \eqref{p3inv} once leads to
\be\label{vteq}
v_z^2=\gm e^{2v}-\de e^{-2v}+2E,
\ee
where $E$  is the same as in \eqref{eq:HamIII} (after transformation). We take the initial conditions $v(\Phi)=\ln a:=b$ and $v'(\Phi)=\frac pa:=q\neq 0$, where $a$ and $p$ are {\em admissible}, and $|\Phi|\gg1$. We decompose the solution into two parts: $v= V+\vh$, where $\vh\ll V$ as $|z|\to\infty$. Substituting into \eqref{vteq}, the equation decouples into 
\begin{subequations}\label{p3v}
\begin{align}
 V''&=\gm e^{2 V}+\de e^{-2 V}\label{p3vcpp},\\
\vh''&=\gm e^{2 V}\sum_{n=1}^{\infty}\frac{(2\vh)^n}{n!}+\frac{\de}{e^{2 V}}\sum_{n=1}^{\infty} \frac{(-2\vh)^n}{n!} -\frac{ V'+\vh'}{z}+\frac{\al e^{ V+\vh}+\beta e^{- V-\vh}}{z},\label{p3vhpp}\\
&=2\vh(\gm e^{2 V}-\de e^{-2 V})-\frac{ V'}{z}+\frac{\al e^{ V}+\beta e^{- V}}{z}\nn\\
&\hspace{2cm}+2\vh^2(\gm e^{2 V}+\de e^{-2 V})-\frac{\vh'}{z}+\vh\frac{\al e^{ V}-\beta e^{- V}}{z}+\ord(z^{-3}).\nn
\end{align}
\end{subequations}
and the initial values become
\ben
 V(\Phi)=b, \; V'(\Phi)=q,\;\vh(\Phi)=0,\; \vh'(\Phi)=0.
\een
Equation \eqref{p3vcpp} tells us that $V=\ord(1)$, and $V$ and $V''$ have the same order. A dominant balance analysis on each leading order term in \eqref{p3vhpp} tells us that $\vh=\ord(1/z)$; $\vh$, $\vh'$ and $\vh''$ are of the same order, moreover $V$ and $V'$ are also of the same order. After integrating equation \eqref{p3vcpp} once, we define two integrals (analytically continued),
\beq
\omt_i&=&\oint_{\Gm_i}\sqrt{\gm e^{2 v}+2E - \de e^{-2 v}}\,d v, \text{ and}\label{omtinv}\\
\om_i&=&\oint_{\Gm_i}\frac{1}{\sqrt{\gm e^{2 v}+2E - \de e^{-2 v} } } \,d v,\label{ominv}
\eeq
where $i=1,2$. These are elliptic integrals when expressed in $u$. The linearly independent contours to be taken are depicted in Fig 1.
\begin{figure}[h!]\label{contourp3}
\begin{tikzpicture}[scale=0.5]
\draw[->] (-5,0)--(6,0) node[right]{};
\draw[->] (-4,-5)--(-4,5) node[above]{};
\draw[fill] (3,1) circle [radius=0.05];
\draw[fill] (-3,-1) circle [radius=0.05];
\draw[fill] (-2,2.5) circle [radius=0.05];
\draw[fill] (2,-2.5) circle [radius=0.05];
\draw[thick, decoration={markings, mark=at position 0.125 with {\arrow[line width=1.4pt]{>}}}, postaction={decorate}] (2.2,-0.2) ellipse (20mm and 35mm);
\node at (4,4) {$\Gm_1$};
\draw[thick, decoration={markings, mark=at position 0.125 with {\arrow[line width=1.4pt]{>}}}, postaction={decorate}] (0.4,2) ellipse (42mm and 18mm);
\node at (4.9,-1) {$\Gm_2$};
\end{tikzpicture}
\caption{The contour choice for $\om$ and $\omt$ integrals.}
\end{figure}
\subsection{Averaging of the Hamiltonian}
In this subsection the averaging of $E_{\rm III}$ defined by $\Delta E:=E(\Phi+\Om_i)-E(\Phi)$, $i=1,2$, is  calculated, where
\ben
E=\frac12\left(v_z^2-\gm e^{2v}+\de e^{2v}\right),
\een
and the subscript $i$ will be dropped hereafter for convenience. Our task is therefore to evaluate
\be\label{p3ave}
E(\Phi+\Om)-E(\Phi)=\frac 1 2 \left(v_z^2(\Phi+\Om)-v_z^2(\Phi)\right).
\ee
In order to find $v_z(\Phi+\Om)$, we first use a Taylor expansion on $v(z)$ around $\Phi+\om$. We remark here that we expect $\Om$ is close to $\om$. Setting $z=\Phi+\Om$ and substituting $v= V+\vh$, we obtain
\ben
v(\Phi+\Om)=\left. \left(V+\vh+(\Om-\om)( V'+\vh')+\frac{(\Om-\om)^2}{2}( V''+\vh'')\right)\right|_{z=\Phi+\om}+\ord((\Om-\om)^3).
\een
where $'=\frac{d}{d\,z}$. We solve this equation for $\Om-\om$ to find
\be\label{p3Om}
\Om-\om=\left.\left(-\frac{\vh}{ V'}+\frac{\vh\vh'}{( V')^2}-\frac{ V''\vh^2}{2( V')^3}\right)\right|_{z=\Phi+\om}+\ord((\Om-\om)^3),
\ee
which shows that $\Om-\om$ is of order $\vh$, therefore is of order $\frac1z$. Using \eqref{p3Om} and performing the analogous expansion on $v_z$, we obtain
\beqn
v_z^2(\Phi+\Om)&=&\left.\left(( V')^2+2( V'\vh'- V''\vh)+(\vh')^2-2\vh\vh''+\frac{ V'''\vh^2}{ V'}\right)\right|_{z=\Phi+\om}+\ord(z^{-3}),
\eeqn
where $ V'''=2 V'(\gm e^{2 V}-\de e^{-2 V})$. Therefore the expression \eqref{p3ave} becomes
\beq
&&E(\Phi+\Om)-E(\Phi)\label{E31}\\
&=&\hspace{-1mm}\left.\left(( V'\vh'- V''\vh)+\frac 1 2 (\vh')^2-\vh\vh''+(\gm e^{2 V}-\de e^{-2 V})\vh^2\right)\right|_{z=\Phi+\om}+\ord(z^{-3}).\nn
\eeq
The goal is to rewrite \eqref{E31} such that the leading order depends only on $V$ and its derivatives. Notice that $\vh$ and $\vh'$ vanish at $z=\Phi$, hence we can rewrite \eqref{E31} as 
\be\label{dE}
\Delta E=\left[\frac{\al e^{ V}-\beta e^{- V}}{z}\right]_{\Phi}^{\Phi+\om}+\int_\Phi^{\Phi+\om}-\frac{( V')^2}{z}-\frac{ V'\vh'- V''\vh}{z}+\frac{\al e^{ V}-\beta e^{- V}}{z^2}+\ord(z^{-3})\,dz.
\ee
Now, using \eqref{p3v}, we find
\ben
\frac{ V'\vh'- V''\vh}{z}=\frac{\al e^{ V}-\beta e^{- V}}{z^2}-\frac{\al e^{ V(\Phi)}-\beta e^{- V(\Phi)}}{\Phi\, z}+\frac 1 z\int_{\Phi}^{z}-\frac{( V')^2}{z}\,dz +\ord(1/z^3).
\een
Using this, we obtain
\beqn
\Delta E\nn&=&\left[\frac{\al e^{ V}-\beta e^{- V}}{z}+\frac{\al e^{ V(\Phi)}-\beta e^{- V(\Phi)}}{\Phi}\ln z \right]_{\Phi}^{\Phi+\om}\nn\\
&&\hspace{15mm}+\int_\Phi^{\Phi+\om}-\frac{( V')^2}{z}+\frac 1 z\int_{\Phi}^{z}\frac{( V')^2}{z}\,dz +\ord(1/z^3)\,dz.
\eeqn
Note that the expansion of $\frac1z$ around $\Phi$ is:
\ben
\frac 1 z=\frac{1}{\Phi+z-\Phi}=\sum_{n=0}^{\infty}\frac{[-(z-\Phi)]^{n+1}}{\Phi^n}=\frac 1 \Phi-\frac{z-\Phi}{\Phi^2}+\frac{(z-\Phi)^2}{\Phi^3}+\ord(1/\Phi^4),
\een
together they yield the result
\beq
\hspace{-15mm}E(\Phi+\Om)-E(\Phi)\nn&=&-\frac{\omt}{\Phi}+\frac{\om\,\omt}{\Phi^2}+\ord(1/\Phi^3),\label{p3E}
\eeq
using integration by parts. This shows that the Hamiltonian varies slowly and only depends on the initial point $\Phi$.
\subsection{The Picard-Fuchs equation for the elliptic integral $\omt$}
In this subsection the Picard-Fuchs equation for $\omt$ is derived, leaving some of the detail to Appendix \ref{B}. Consider the indefinite incomplete elliptic integral
\beqn
\phit(z)&:=&\int_{ v(\Phi)}^{ v(z)}\sqrt{\gm e^{2 v}+2E-\de e^{-2 v}}d v, \\
\phi(z)&:=&\int_{ v(\Phi)}^{ v(z)}\frac{1}{\sqrt{\gm e^{2 v}+2E-\de e^{-2 v}}}d v.
\eeqn
Note that $\omt=\phit(\Phi+\om)$ and $\om=\phi(\Phi+\om)$.  Integrating $\phit$ by parts yields the alternative form  (see Appendix \ref{B}):
\be\label{phiteq}
\phit= v_z(z)- v_z(\Phi)+2E\phit'-2\de\Psi
\ee
where $'=\frac{d}{dE}$, and
\beqn
\Psi&:=&\int_{ v(\Phi)}^{ v(z)}\frac{1}{e^{ v}\sqrt{\gm e^{4 v}+ 2E e^{2 v}-\de}} d v,
\eeqn
and a further computation will show that (see Appendix \ref{B})
\beqn
\Psi=\frac{1}{e^{2 v(\Phi)} v_z(\Phi)}-\frac{1}{e^{2 v(z)}v_z(z)}+ 2\gm \phit''+2 E \Psi'.
\eeqn
Eliminating $\Psi$ and its derivatives from \eqref{phiteq} and its derivative in $E$ leads to the following second-order differential equation satisfied by $\phit$:
\be \label{pfphip3}
4(E^2+\de\gm)\phit''+\phit- \left.\frac{\gm e^{2 v}+\de e^{-2 v}}{ v_z}\right|_{ v(\Phi)}^{ v(z)}=0.
\ee
In the special case that $\phit(\Phi+\om)=\omt$, the Picard-Fuchs equation satisfied by $\omt$ is obtained:
 \ben
4(E^2+\de\gm) \omt''+\omt=0.
\een
We note that the singularities of the above equation occur where the elliptic function $ U$ becomes degenerate.
 \subsection{The expansion of distance $\Om$ to the next repeated value}
Equation \eqref{p3Om} can be understood as an expansion of $\Om$ in $v$ in which the leading order term is simply $\om$. The expression $\vh/ V_z$ is evaluated in Appendix \ref{B}. The distances to the next repeated value of the initial value in $u$ for $\pIII$ have the expansion
\beqn
\Om&=&\om-\frac{\omt\om'}{2\Phi}-\frac{1}{\Phi}\left[\frac{\al u^2(\Phi)-\beta}{ u(\Phi)}-\frac{\gm u^4(\Phi)+\de}{ u(\Phi) u_z(\Phi)}\right]+\ord(1/\Phi^2).\label{Omexpp3}
\eeqn
This concludes our calculation for $\pIII$.
 \subsection{Boundedness of $E$}\label{bound}
In this section we show that $E_{III}$ is bounded along a ray with fixed angle. We expect that the governing equation for $E$ are the same as $\pI$ and $\pII$ to leading order since $E_{\rm I}-E_{\rm V}$ possess the same behaviour to leading order. $E(z)$, which may not be analytic in $z$ because its derivative is path dependent, see \cite{JoshiThesis}, have the following `Taylor' expansion around $\Phi$ using Wirtinger derivative,
\beqn
E(z)&=& E+(z-\Phi)E_z+(\zb-\bar{\Phi})E_{\zb}\\
&&\hspace{-5mm}+\frac12\left[(z-\Phi)^2E_{zz}+(z-\Phi)(\zb-\Phib)E_{z\zb}+(\zb-\Phib)^2E_{\zb\zb}\right]+\cdots
\eeqn
evaluating at $(\Phi+\Om_j,\Phib+\Omb_j)$ gives us
\beq
E(\Phi+\Om_j)&=& E(\Phi)+\Om_j E_z+\Omb_j E_{\zb}+\frac12\left[\Om_j^2 E_{zz}+\Om_j \Omb_j E_{z\zb}+\Omb_j ^2E_{\zb\zb}\right]+\ord\left(\frac{1}{z^2}\right).\nn
\eeq
After expanding $\Om$ by using \eqref{Omexpp3}, this becomes
\beq
E(\Phi+\Om_j)- E(\Phi)&=&\om_j E_z+\omb_j E_{\zb}+\frac12\left[\om_j^2 E_{zz}+\om_j \omb_j E_{z\zb}+\omb_j ^2E_{\zb\zb}\right]+\ord\left(\frac{1}{z^2}\right)\nn\\
&=&-\frac{\omt}{\Phi}+\frac{\om\,\omt}{\Phi^2}+\ord(\frac{1}{\Phi^3})\nn\\
\Ra&&\om_j E_z+\omb_j E_{\zb}=-\frac{\omt_j}{z}+\ord(\frac{1}{z^2})\label{Ediff}.
\eeq
Converting \eqref{Ediff} to polar coordinates $z=r e^{i\tht}$, then applying $t=\ln r$ gives us
\be
E_t=\frac{e^{-2i\tht} \ka-Y}{Y_E}+\ord(e^{-t}).
\ee
where
\beqn
\ka&:=&\omt_1\om_2-\om_1\omt_2,\\
Y&:=&\omt_1\omb_2-\omb_1\omt_2,\\
Y_E&:=&\om_1\omb_2-\omb_1\om_2.
\eeqn
Using results from Appendix \ref{C}, we obtain
\beqn
E_t=-2E+2\frac{e^{-2i\tht}|E|+E}{\ln\left(\frac{8|E|}{|\mu|^{1/2}}\right)}+\ord\left(\frac1E,\frac{1}{E\ln|E|}\right)
\eeqn
where $\mu=-\gm\de$. This proves that $E$ is bounded on a path to infinity at any fixed angle.
\section{The fourth \pv equation}\label{sec:p4}
In this section we briefly repeat the analysis described in Section \ref{sec:p3}  for the fourth {\pv} equation \eqref{p4u}. The designation $J=\rm IV$ and the subscript $i$ will be dropped for convenience. Starting with the transformation $u=\frac14 v^2$, equation \eqref{p4u} is converted to
\be
v_{zz}=\frac{3}{64} v^5+\frac{v^3}{2}+v-\frac{v_z}{z}-\frac{\al v}{2z}+\frac{2\beta}{z^2 v^3}+\frac{v}{16z^2},\label{transp4}
\ee
which does not have a first derivative term in the leading order. This equation has the corresponding Hamiltonian,
\be
E_{\rm IV}=\frac12\left(v_z^2-\frac{v^6}{64}-\frac{v^4}{4}-v^2\right).
\ee
The initial conditions are now $v(\Phi)=b$ and $v'(\Phi)=q\neq0$ where $|\Phi|\gg1$. The decomposition $v=V+\vh$, where $\vh\ll V$, decouples equation \eqref{transp4} into:
\beq
 V_{zz}&=&\frac 3 {64} V^5+\frac12 V^3+ V\label{p4vceq},\\
\vh_{zz}&=&\vh\left(\frac{15}{64} V^4+\frac32 V^2+1\right)-\frac{2 V_z+\al  V}{2z}\nn\\
&&+\vh^2\left(\frac{15}{32} V^3+\frac32 V\right)-\frac{2\vh_z+\al \vh}{2z}+\frac{ V}{16z^2}+\frac{2\beta}{z^2 V^3}\nn+\ord(1/z^3),\nn
\eeq
and the initial conditions become
\ben
V(\Phi)=a,\;V'(\Phi)=q,\; \vh(\Phi)=0,\;\vh'(\Phi)=0.
\een
By using the same argument as before we find that $ V=\ord(1)$ and $\vh=\ord(1/z)$; $\vh$, $\vh'$ and $\vh''$ are of the same order; and $V$, $V'$ and $V''$ are of the same order. 
The change in the Hamiltonian is
\beqn
&&E(\Phi+\Om)-E(\Phi)\\
&=&\left. \left(V'\vh'- V''\vh+\frac12(\vh')^2 - \vh\vh'' +\frac12\left(\frac{15}{64} V^4+\frac32 V^2+1\right)\vh^2\right)\right|_{z={\Phi+\om}}+ \ord(z^{-3})\nn\\
&=&-\frac{\omt}{\Phi}+\frac{\om\omt}{\Phi^2}+\ord(\Phi^{-3})
\eeqn
As before, two incomplete elliptic integrals $\phi$ and $\phit$ are defined from \eqref{p4vceq}, and  from using \eqref{periodints},
\beqn
\phi(z)&=&\int_{ v(\Phi)}^{ v(z)}\frac{d v}{\sqrt{\frac{ v^6}{64}+\frac{ v^4}{4}+ v^2+2E}},\\
\phit(z)&=&\int_{ v(\Phi)}^{ v(z)}\sqrt{\frac{ v^6}{64}+\frac{ v^4}{4}+ v^2+2E} d v.
\eeqn
The integral $\phit$ can be expressed as 
\ben
\phit=\frac14  v v_z|_{ v(\Phi)}^{ v(z)}+\frac{3}{2}E\phit'+\frac1{16}\Psi,
\een
where $'=\frac{d}{dE}$, and
\beqn
\Psi&:=&\int_{ v(\Phi)}^{ v(z)}\frac{v^4+8v^2}{\sqrt{\frac{ v^6}{64}+\frac{ v^4}{4}+ v^2+2E_0}}d v,\\
&=&\left.\frac{(\frac{1}{2} v^5+\frac{20}{3} v^3)}{ v_z}\right|_{ v(\Phi)}^{ v(z)}-\frac{128}{3}\phit'-\left(\frac{16}{3}-3E\right)\Psi'-\frac{256E}{3}\phit''.
\eeqn
Together we have
\ben
\left(\frac92 E-\frac83\right)E\phit''+\phit-\left[\frac{ v v_z}{4}+\frac{3 v^5 + 40 v^3+8(16-9 E)v}{ 96v_z}\right]_{ v(\Phi)}^{ v(z)}=0.
\een
At $\phit(\Phi+\om)=\omt$, we have the Picard-Fuchs equation for $\omt$
\ben
\left(\frac92 E-\frac83\right)E\,\omt''+\omt=0.
\een	
The distance to the next repeated initial value is:
\begin{align}
\Om(\Phi)&=\om-\frac{\omt\om'}{2\Phi}+\frac{\om'}{\Phi}\Bigl(\frac{3 v^7(\Phi) + 72 v^5(\Phi) + 512 v^3(\Phi) + 64(16-3E) v(\Phi)}{768 v_z(\Phi)}\label{Om4v}\\
&\phantom{=\om-\frac{\omt\om'}{2\Phi}+\frac{\om'}{\Phi}\Bigl(} 
\ +\frac{\al v^2(\Phi)}{4} \Bigr)+\ord\left(\frac{1}{\Phi^2}\right).\nn
\end{align}
After transforming equation \eqref{Om4v} back to the variable $u$, we have the result stated in Theorem \ref{thm}. Following the argument from Section \ref{bound} and using results from Appendix \ref{C} give us
\be
E_t=-\frac32 i E+\frac{3a_1}{2a_0}e^{-2i\tht}|E|^{2/3}+\frac{3 a_1^2E^{2/3}}{2a_0^2\Eb^{1/3}}+\ord(1),
\ee
which shows that $E$ is bounded along a ray going to infinity.
\section{The fifth \pv equation}\label{sec:p5}
As in the previous section, we briefly repeat the analysis for equation \eqref{p5u}, with the understanding that $\rm J=\rm V$ and $i$ will remain unspecified. The transformation $u = \left(\frac{e^v+1}{e^v-1}\right)^2$ eliminates the first derivative terms in the leading order part of $\pV$, obtaining
\be
v_{zz}=-\frac{\de}{8e^{2v}}\left(e^{4v}-1\right)-\frac{\gm(e^{2 v}-1)}{4 e^v z}-\frac{v_z}{z}-\frac{4 e^{v} \left[\al(e^v+1)^4+\beta (e^v-1)^4\right]}{z^2 \left(e^{2v}-1\right)^3}.\label{pvbbdep}
\ee
The Hamiltonian of $\pV$ transforms into the following form
\be
E=\frac12\left(v_z^2+\frac{\de}8 (e^{2v}+e^{-2v})-\frac{\de}{4}\right),
\ee
the term $\frac{\de}{4}$ is added for consistency with the $u$-equation. The initial conditions are taken to be $v(\Phi)=b$ and $v'(\Phi)=q\neq0$ where $|\Phi| \gg1$. Analogously we use the decomposition $v=V+\vh$, where $\vh\ll V$ as $|z|\to\infty$, the equation \eqref{pvbbdep} and the initial conditions decouple into
\beqn
 V_{zz}&=&-\frac{\de}{8}(e^{2 V}-e^{-2 V}),\label{p5vcpp}\\
\vh_{zz}&=&-\frac{\de}{4} (e^{2 V}-e^{-2 V}) \vh-\frac{\gm}{4z}(e^{ V}-e^{- V})-\frac{ V'}{z}-\frac{\de}{4}(e^{2 V}-e^{-2 V})\vh^2\nn\\
&&-\frac{\gm e^{ V}}{4z}\vh(e^{ V}+e^{- V})-\frac{\vh'}{z}-\frac{4 \left[\al  (e^{ V}+1)^4+\beta (e^{ V}-1)^4\right]}{z^2}+\ord(1/z^3),\nn
 \eeqn
and
\ben
 V(\Phi)=a,\;V'(\Phi)=q,\;\vh(\Phi)=0,\;\vh'(\Phi)=0.
\een
By using another dominant balance analysis we find that $ V=\ord(1)$ and $\vh=\ord(1/z)$; $\vh$, $\vh'$ and $\vh''$ are of the same order; and $V$, $V'$ and $V''$ are of the same order. The evolution of the Hamiltonian $E$ is
\beqn
&&E(\Phi+\Om)-E(\Phi)\\
&=&\left. \left(V'\vh'- V''\vh+\frac12(\vh')^2-\vh\vh''-\frac{\vh^2\de}{8}(e^{2 V}+e^{-2 V})\right)\right|_{z=\Phi+\om}+\ord(z^{-3})\\
&=&-\frac{\omt}{\Phi}+\frac{\om\omt}{\Phi^2}+\ord(\Phi^{-3}).
\eeqn
The following integrals are defined corresponding to the analytically continued elliptic integrals for $v$ using \eqref{periodints}:
\beqn
\phi(z)&=&\int_{ v(\Phi)}^{ v(z)}\frac{1}{\sqrt{-\frac{\de}{8}(e^{2 v}+e^{-2 v})+2E+\frac{\de}{4}}}\,d v,\\
\phit(z)&=&\int_{ v(\Phi)}^{ v(z)}\sqrt{-\frac{\de}{8}(e^{2 v}+e^{-2 v})+2E+\frac{\de}{4}}d v.\\
\eeqn
The differential equations for $\phit$ is,
\ben
\left(4E^2-\frac{\de^2}{16}\right)\phit''+\phit+\left.\frac{\de}{8}\frac{e^{2 v}-e^{-2 v}}{ v_z}\right|_{ v(\Phi)}^{ v(z)}=0,
\een
and at $\phit(\Phi+\om)=\omt$, the above equation becomes
\ben
\left(4E^2-\frac{\de^2}{16}\right)\omt''+\omt=0.
\een
The exact distance to the next repeated value of $u(\Phi)$ of the fifth \pv equation takes the form
\beqn
\Om(\Phi)&=&\om-\frac{\omt\om'}{2\Phi}-\frac{\om'}{\Phi}\left[-\frac{\gm}{4}(e^{ v(\Phi)}+e^{- v(\Phi)})+\frac{\de}{8}\frac{e^{-2 v(\Phi)}-e^{2 v(\Phi)}}{ v_z(\Phi)}\right]+\ord(1/\Phi^2).
\eeqn
Once again, transforming the above expression back to the variable $u$ yields the result stated in Theorem \ref{thm}. Following the same argument from Section \ref{bound} and Appendix \ref{C} gives us
\be
E_t=-2E+2\frac{E+|E|e^{-2i\tht}}{\ln\left(\frac{64|E|}{|\de|}\right)}+\ord(1),
\ee
which shows that $E$ is bounded along a ray going to infinity.

\section{Conclusion}
In this paper we showed how to describe the leading-order asymptotic behaviours of the Painlev\'e equations $\pI-\pV$ as the independent variable in each equation becomes large. The resulting behaviours are given by elliptic functions to leading-order in the complex plane. Each elliptic function has a modulus which is related to the Hamiltonian of the respective equation. Our main result shows how the Hamiltonian is bounded, and changes locally as the independent variable moves over a domain given by the local period parallelogram. The Picard-Fuchs equation was derived for each case, and used to estimate intervals between repeated values of the solution. The results expressing the change of each Hamiltonian are remarkably similar for all the equations, despite their differences in form.

The similarity of our results for $\pI-\pV$ suggest a deep connection to the geometry of their initial-value space \cite{O79}. Another open question concerns the discrete Painlev\'e equations which share a similar geometry of initial-value spaces. In the collection of all known discrete Painlev\'e equations only a few cases have been studied asymptotically, by different approaches.

Another intriguing question arises for the case of $\pVI$. Known asymptotic results for $\pVI$ do not appear to involve elliptic functions. On the other hand, our results suggest that some connection to elliptic functions should appear, because $\pVI$ has a coalescence limit to $\pV$. What these behaviours are for $\pVI$ remains an open question.

\appendix
\section{Transformations needed for this Paper}
\subsection{Boutroux Transformations for $\pI$, $\pII$, $\pIV$}\label{A}
The first, second and the fourth \pv equations have the following standard form
\beqn
\pI : && y_{tt}=6y^2+t,\\
\pII :&&y_{tt}=2y^3+ty+\al_0, \text{ and}\\
\pIV :&&y_{tt}=\frac12\frac{y_t^2}{y}+\frac{\beta}{y}+2(t^2-\al)y+4ty^2+\frac32y^3.
\eeqn
Using the Boutroux transformations:
\beqn
y&=&\tau^{1/2}u(z),\quad\text{where }z=\frac45 \tau^{5/4}\quad (\text{where }\tau=-t),\\
y&=&\tau^{1/2}u(z),\quad\text{where }z=\frac23 \tau^{3/2}\quad (\text{where }\tau=-t), \text{ and }\\
y&=&t\,u(z),\quad\text{where }z=\frac12t^2,
\eeqn
after renaming $\al_0=-\frac23\al$, $\pI$, $\pII$ and $\pIV$ are transformed into the Boutroux form \eqref{p1u}, \eqref{p2u} and \eqref{p4u}.
\subsection{Transformation that eliminates the first derivative term in $\pIII-\pV$}\label{A2}
For \eqref{p3u}-\eqref{p5u}, we would like to eliminate the term $\frac{u_z^2}{u}$. One way of doing this is to introduce a new dependent variable $v$ and transformation factor $f$ such that
\ben
f_{\rm III}(u)\left(u_{zz} - \frac{u_z^2}{u}\right)=v_{zz}.
\een
Choosing $f=\frac1u$ corresponds to the simple transformation $v=\ln u$. Similarly for $\pVI$ and $\pV$, we have
\beqn
&&f_{\rm VI}(u)\left(u_{zz} - \frac{u_z^2}{2u}\right)=v_{zz},\\
&&f_{\rm V}(u)\left(u_{zz} - \left(\frac{1}{2u}+\frac{1}{u-1}\right)u_z^2\right)=v_{zz},
\eeqn
where $f_{\rm VI}(u)=\frac{1}{\sqrt u}$, and $f_{\rm V}(u)=\frac{1}{\sqrt u(u-1)}$ which gives us $v=2 u^{1/2}$ and $v=\ln\left(\frac{1-\sqrt u}{1+\sqrt u}\right)$ for $\pIV$, and $\pV$ respectively.

\section{Calculation needed for analysis in Section \ref{sec:p3}}\label{B}
\subsection{The differential equation satisfied by the elliptic integrals $\phit_i$}
In this appendix we provide the details necessary to derive the differential equation satisfied by $\phit$ with respect to $E$. We use a prime to denote differentiation with respect to $E$, and a subscript for differetiation with respect to $z$. To start, we use integration by parts
\beqn
\phit&=& \int_{V (\Phi)}^{V (z)}e^{-V }\sqrt{\gm e^{4V }+2E_0e^{2V }-\de}dV ,\\
&=&-V _z|_{V (\Phi)}^{V (z)}+2\int_{V (\Phi)}^{V (z)}\frac{\gm e^{2V }+2E_0 e^{2V }- \de -E_0 e^{2V }+\de}{e^{V }\sqrt{\gm e^{2V }+E-\de e^{-2V }}}dV ,\\
&=&-V _z|_{V (\Phi)}^{V (z)}+2\phit-2E_0\phi+2\de\Psi.
\eeqn
This allows us to solve for $\phit$ and get
\ben
\phit=V _z|_{V (\Phi)}^{V (z)}+2E_0\phit'-2\de\Psi,
\een
where 
\beqn
\Psi&:=&\int_{ v(\Phi)}^{ v(z)}\frac{1}{e^{ v}\sqrt{\gm e^{4 v}+ 2E e^{2 v}-\de}} d v.
\eeqn
We again use integration by parts to express $\Psi$ in terms of $\phit$, $\Psi$ and their derivatives with respect to $E$:
\beqn
\Psi=\left.-\frac{1}{e^{2 v} v_z}\right|_{v(z)}^{v(\Phi)}+ 2\gm \phit''+2 E \Psi'
\eeqn
where
\beqn
\phit''=\frac{d\phi}{dE}&=&-\int_{ v(\Phi)}^{ v(z)}\frac{e^{3 v}}{(\gm e^{4 v}+2E e^{2 v}-\de)^{3/2}}d v,\\
\Psi'\,=\,\frac{d \Psi}{dE}&=&-\int_{ v(\Phi)}^{ v(z)}\frac{e^{ v}}{(\gm e^{4 v} + 2 E e^{2 v}-\de)^{3/2}}d v.
\eeqn
Putting the above together gives equation \eqref{pfphip3} in which we have also used the readily proven relation $\frac{dv_z}{dE}=\frac{1}{ v_z}$.
\subsection{The first order term in the expansion of $\Om$ for $\pIII$}
In this appendix we calculate the order of $\frac 1z$ term of the expansion of $\Om$ (as defined in \eqref{p3Om}):
\beq
\left.\frac{\vh}{ V_z}\right|_{z=\Phi+\om}&=&\int_{\Phi}^{\Phi+\om}\frac{ V_z\vh_z-\vh V_{zz}}{( V_z)^2}dz+\ord(\Phi^{-2}),\nn\\
&=&\int_{\Phi}^{\Phi+\om}\frac{1}{( V_z)^2}\left[\frac{\al e^{ V}-\beta e^{- V}}{z}\right]_{\Phi}^z-\frac{1}{( V_z)^2}\int_{\Phi}^{z}\frac{( V_z)^2}{z}dz\, dz+\ord(\Phi^{-2}),\nn\\
&=&\frac{1}{\Phi}\oint\frac{\al e^{ V}-\beta e^{- V}}{( V_z)^3}d V-\frac{\al e^{ V(\Phi)}-\beta e^{- V(\Phi)}}{\Phi}\oint \frac{1}{( V_z)^3}d V\nn\\
&&\hspace{25mm} -\frac{1}{\Phi}\oint\frac{1}{( V_z)^3}\int_{ V(\Phi)}^{ V(z)} V_zd V\, d V+\ord(\Phi^{-2})\label{123intp3}.
\eeq
The first integral in \eqref{123intp3} is
\beqn
\oint_C\frac{\al e^{ V}-\beta e^{- V}}{( V_z)^3}d V&=&\oint_C\frac{\al e^{4 V}-\beta e^{2 V}}{(\gm e^{4 V}+2E_0 e^{2 V}-\de)^{3/2}}d V,\\
&=&\oint_{\Gm}\frac{\al  U^3-\beta U}{(\gm  U^4+2 E_0  U^2-\de)}d U,\\
&=&\oint_{\tilde{\Gm}}\frac{\al\wc-\beta}{2(\gm\wc^2+2E_0\wc-\de)^{3/2}}d\wc.
\eeqn
In the variable $ U$, we can see clearly that the four branch points are $ U_{\pm\pm}=e^{ V}_{\pm\pm}=\pm\sqrt{\frac{-E_0\pm\sqrt{E_0^2+\gm\de}}{\gm}}$ and that each will generically occupy its own quadrant. If the contour $\tilde{\Gm}$ circles two branch points of the opposite sign, then the integral is zero by cancellation. If the contour $\tilde{\Gm}$ circles any other two branch points, then in the variable $\wc$ the contour encloses both of the branch points, so we can enlarge the contour and instead calculate the residue at infinity. The residue is indeed $0$ in this case. The second integral is simply $\frac{d\om}{dE_0}$. For the third integral in \eqref{123intp3}
\beqn
&&\oint\frac{1}{( V_z)^3}\int_{ V(\Phi)}^{ V(z)} V_z d V\, d V, \\
& = & \oint \frac{1}{( V_z)^3} \phit\, d V,\\
& = & \oint 4(E_0^2+\de\gm)\phi_E \pa_{ V}\phi_E d V-\oint \frac{1}{( V_z)^3} \left[\frac{\gm e^{2 V(\Phi)}+\de e^{-2 V(\Phi)}}{ V_z(\Phi)}\right]\, d V\\
&&\phantom{\oint 4(E_0^2+\de\gm)\phi_E\pa_{ V}\phi_Ed V}  +\oint \frac{1}{( V_z)^3}\left[\frac{\gm e^{2 V}+\de e^{-2 V}}{ V_z}\right]\, d V,\\
&=&4(E_0^2+\de\gm)\frac{1}{2}\left(\frac{d\,\om}{dE_0}\right)^2+\frac{\gm e^{2 V(\Phi)}+\de e^{-2 V(\Phi)}}{ V_z(\Phi)}\frac{d\,\om}{dE_0},\\
&=&-\frac{\omt}{2}\frac{d\,\om}{dE_0}+\frac{d\,\om}{dE_0}\frac{\gm e^{2 V(\Phi)}+\de e^{-2 V(\Phi)}}{ V_z(\Phi)}.
\eeqn
Putting this together, we have the desired result  
\ben
\frac{\vh}{V_z}=\frac{\omt\om'}{2\Phi}+\left[\al e^{ V(\Phi)}-\beta e^{- V(\Phi)}-\frac{\gm e^{2 V(\Phi)}+\de e^{-2 V(\Phi)}}{ V_z(\Phi)}\right]\frac{\om'}{\Phi}+\ord(1/\Phi^2).
\een
\section{Expansions of $\omt$ in $E$ as $E\to\infty$}\label{C}
For $\pIII$, 
\beqn
\omt_1&=&2\sqrt2\pi i E^{1/2}w_0,\\
\omt_2&=&\sqrt2\left(\ln\frac\mu4-4(\ln2-1)-2\ln E\right)E^{1/2}w_0+E^{1/2} w_1,
\eeqn
where
\beqn
w_0&=&1-\frac{\mu}{16}\frac{1}{E^2}-\frac{15\mu^2}{1024}\frac{1}{E^4}+\ord(E^{-6}),\\
w_1&=&\frac{\mu}{16}\frac{1}{E^2}+\frac{13\,\mu^2}{2048}\frac{1}{E^4}+\ord(E^{-6}).
\eeqn
For $\pIV$:
\beqn
\omt_1&=&\frac{3(3-\sqrt{3}i)}{4}a_0E^{2/3}w_0+\frac{3(3+\sqrt{3}i)}{2}a_1E^{1/3}w_1,\\
\omt_2&=&\frac{3(3+\sqrt{3}i)}{4}a_0E^{2/3}w_0+\frac{3(3-\sqrt{3}i)}{2}a_1E^{1/3}w_1,
\eeqn
where
\beqn
w_0&=&1-\frac{16}{81}\frac1E-\frac{512}{32805}\frac1{E^2}+\ord(E^{-3}),\\
w_1&=&1-\frac{8}{81}\frac1E-\frac{640}{45927}\frac1{E^2}+\ord(E^{-3}),\\
a_0&=&\int_{0}^{\infty}\frac{1}{\sqrt{v^4+2v}}dv,\\
a_1&=&\int_{0}^{\infty}-\frac{2v^{3/2}}{(v^3+2)^{3/2}}dv.
\eeqn
For $\pV$:
\beqn
\omt_1&=&-\sqrt2\left[-w_1E^{1/2}\ln(E)+\frac{1}{E^{1/2}}w_0\right]+\left(\ln\left(\frac{64}{\de}\right)-2\right)\sqrt2 E^{1/2}w_1,\\
\omt_2&=&2\sqrt2\pi\,i\,E^{1/2}w_1,
\eeqn
where
\beqn
w_0&=&-\frac{\de}{8}+\frac{\de^2}{1024}\frac1E+\frac{\de^3}{49152}\frac{1}{E^2}+\ord(E^{-3}),\\
w_1&=&1+\frac{\de}{16}\frac1E-\frac{3\,\de^2}{1024}\frac{1}{E^2}+\frac{5\,\de^3}{16384}\frac{1}{E^3}+\ord(E^{-4}).
\eeqn

\bibliographystyle{plain}

\end{document}